\input{aipcheck}

\documentclass[
    ,final            
  ]
  {aipproc}

\layoutstyle{6x9}

\def \GeV{{\mathrm{GeV}}}

\newcommand{\mdm}{{m_{DM}}}

\newcommand{\Rsun}{{R_{\odot}}}

\def\lsim{\raise0.3ex\hbox{$\;<$\kern-0.75em\raise-1.1ex\hbox{$\sim\;$}}}
\def\gsim{\raise0.3ex\hbox{$\;>$\kern-0.75em\raise-1.1ex\hbox{$\sim\;$}}}

\def    \bea           	{\begin{eqnarray}}
\def    \eea           	{\end{eqnarray}}

\newcommand{\gev}{\,\textrm{GeV}}

\newcommand{\cm}{\,\textrm{cm}}

\newcommand{\photino}{\tilde{\gamma}}

\newcommand{\taudm}{{\tau_{DM}}}

\begin{document}

\title{The $\mu\nu$SSM and gravitino dark matter}


\author{Carlos Mu\~noz}{  
address={Departamento de F\'{\i}sica Te\'{o}rica and Instituto de F\'{\i}sica Te\'{o}rica UAM/CSIC,\\ Universidad Aut\'{o}noma de Madrid,
Cantoblanco, E-28049 Madrid, Spain}
}



\begin{abstract}
We consider the phenomenological implications of gravitino dark matter in the context of the $\mu\nu$SSM. 
The latter is an R-parity breaking model which provides a solution to the
$\mu$-problem of the MSSM and explains the origin of neutrino masses
by simply using right-handed neutrino superfields. 
In particular, we analyze the prospects for detecting gamma rays 
from decaying gravitinos.
Gravitino masses larger than $20$ GeV are disfavored by 
the isotropic diffuse photon background measurements, but
a gravitino with a mass range between $0.1 - 20$ GeV gives rise to 
a signal that might easily be observed by the FERMI satellite. 
Through this kind of analysis important regions of the parameter space of the $\mu\nu$SSM can be
checked.

\end{abstract}

\maketitle


\section{Introduction}

Supersymmetry (SUSY) is still one of the most attractive
theories for physics beyond the Standard Model, and we expect to find its
signatures in the forthcoming LHC. However, SUSY has also theoretical problems, and, in particular, a very important one is the so-called
$\mu$-problem.
This problem arises from the requirement of a SUSY mass term for the Higgs fields
in the superpotential of the Minimal Supersymmetric Standard Model (MSSM),
$\mu \hat H_u \hat H_d$, which must be of the order of the
electroweak scale to successfully lead to electroweak symmetry breaking.
In the presence of a GUT and/or a gravitational theory with typical scales $10^{16}$ and $10^{19}$ GeV, respectively, one should explain how to obtain a SUSY mass term of the order of $10^{2}-10^{3}$ GeV.

On the other hand, neutrino experiments have confirmed during the last years that
neutrinos are massive. As a consequence, all theoretical models must be modified in order to reproduce this result.

The ``$\mu$ from $\nu$'' Supersymmetric Standard Model
($\mu\nu$SSM) 
was proposed in the literature \cite{MuNuSSM,MuNuSSM0}\footnote{
Several recent papers have studied different aspects of the $\mu\nu$SSM.
See the works in \cite{MuNuSSM2,Ghosh:2008yh,Hirsch0,neutrinos}.
}
as an alternative to the MSSM.
In particular, it provides a solution to the
$\mu$-problem and explains the origin of neutrino masses
by simply using right-handed neutrino superfields.

The superpotential of the $\mu\nu$SSM contains, in addition to the
usual Yukawas for quarks and charged leptons,
Yukawas for neutrinos
$\hat H_u\,  \hat L \, \hat \nu^c$, terms of the type
$\hat \nu^c \hat H_d\hat H_u$ 
producing an effective  $\mu$ term through right-handed sneutrino
vacuum expectation values (VEVs),
and also terms of the type $\hat \nu^c \hat \nu^c \hat \nu^c$  
avoiding the existence of a Goldstone boson and
contributing to generate
effective Majorana masses for neutrinos at the electroweak scale.
Actually, the explicit breaking of R-parity in this model 
by the above terms produces the mixing of neutralinos with
left- and right-handed neutrinos, and as a consequence a generalized matrix of the
seesaw type that gives rise at tree level to three
light eigenvalues corresponding to neutrino 
masses \cite{MuNuSSM}.
It is worth noticing here that this possibility of using a seesaw at the
electroweak scale avoids the introduction of {\it ad-hoc} high energy scales
in the model.


The breaking of R-parity can easily be understood if we realize that in the limit 
where neutrino Yukawa couplings are vanishing, the 
$\hat \nu^c$ are just ordinary singlet superfields, 
without any connection with neutrinos,
and
this model would coincide (although with three instead of one singlet) with the
Next-to-Minimal Supersymmetric Standard Model (NMSSM) where R-parity is 
conserved.
Once we switch on the neutrino Yukawas, the fields
$\hat \nu^c$ become right-handed neutrinos, and, as a consequence, R-parity
is broken. Indeed this breaking is small because, as mentioned above, we have an electroweak scale seesaw, implying neutrino Yukawas no larger than $10^{-6}$ (like
the electron Yukawa) to reproduce the neutrino masses ($\lsim 10^{-2}$ eV).

Since R-parity is broken, one could worry about fast proton decay through the usual baryon and lepton number violating operators of the MSSM.
Nevertheless, the choice of $R$-parity is {\it ad hoc}. There are other discrete 
symmetries, 
like e.g. baryon triality which only forbids the baryon violating operators \cite{dreiner3}.
Obviously, for all these symmetries R-parity is violated.
Besides, in string constructions the matter superfields can be located in different sectors of the compact space or have different extra $U(1)$ charges, in such a way that 
some operators violating $R$-parity can be forbidden \cite{old}, 
but others can be allowed.

On the other hand, 
when $R$-parity is broken the lightest supersymmetric particle (LSP) is no longer stable. Thus
neutralinos 
or sneutrinos, 
with
very short lifetimes,
are no longer candidates for the dark matter (DM) of the Universe.
Nevertheless, if the gravitino is the LSP its decay is suppressed both by
the gravitational interaction and by the small R-parity violating coupling, and as a 
consequence its lifetime can be much longer than the age of the 
Universe \cite{Takayama:2000uz}.
Thus the gravitino can be in principle a DM candidate in R-parity breaking models.
This possibility and its phenomenological consequences were studied mainly in the context of bilinear or trilinear R-parity violation 
scenarios in \cite{Takayama:2000uz,Hirsch,buchmuller,Lola,bertone,ibarra,tran,moro,covi}.
In \cite{buchmuller,bertone,ibarra,moro} the prospects for detecting gamma rays
from decaying gravitinos
in satellite experiments were also analyzed.
In a recent work \cite{recentgravitino} we have discussed these issues, gravitino 
DM and
its possible detection in the FERMI satellite, in the context of the $\mu\nu$SSM.
In this talk I will summarize the results obtained in this work.


\section{The $\mu\nu$SSM}

The  superpotential of the $\mu \nu$SSM introduced in \cite{MuNuSSM} is 
given by
\begin{eqnarray}
W & = &
\ \epsilon_{ab} \left(
Y_{u_{ij}} \, \hat H_u^b\, \hat Q^a_i \, \hat u_j^c +
Y_{d_{ij}} \, \hat H_d^a\, \hat Q^b_i \, \hat d_j^c +
Y_{e_{ij}} \, \hat H_d^a\, \hat L^b_i \, \hat e_j^c +
Y_{\nu_{ij}} \, \hat H_u^b\, \hat L^a_i \, \hat \nu^c_j 
\right)
\nonumber\\
& - &
\epsilon{_{ab}} \lambda_{i} \, \hat \nu^c_i\,\hat H_d^a \hat H_u^b
+
\frac{1}{3}
\kappa{_{ijk}} 
\hat \nu^c_i\hat \nu^c_j\hat \nu^c_k \ .
\label{superpotential}
\end{eqnarray}
%
In addition to terms from 
$L_{soft}$, the 
tree-level scalar potential receives the $D$ and $F$ term
contributions. The final neutral scalar potential can be found 
in \cite{MuNuSSM, MuNuSSM2}.
In the following we will assume for simplicity that all parameters in the potential are real. 
Once the electroweak symmetry is spontaneously broken, the neutral scalars develop in general the following VEVs:
\begin{equation}\label{2:vevs}
\langle H_d^0 \rangle = v_d \, , \quad
\langle H_u^0 \rangle = v_u \, , \quad
\langle \tilde \nu_i \rangle = \nu_i \, , \quad
\langle \tilde \nu_i^c \rangle = \nu_i^c \,.
\end{equation}

For our computation below we are interested in the neutral fermion mass matrix.
As explained in \cite{MuNuSSM,MuNuSSM2},
neutralinos mix with the neutrinos and therefore in a basis where
${\chi^{0}}^T=(\tilde{B^{0}},\tilde{W^{0}},\tilde{H_{d}},
\tilde{H_{u}},\nu_{R_i},\nu_{L_i})$,
one obtains the following neutral fermion mass terms in the Lagrangian
$-\frac{1}{2} (\chi^0)^T \mathcal{M}_{\mathrm{n}}\chi^0 + \mathrm{c.c.}$,
where
\begin{equation}
{\mathcal{M}}_n=\left(\begin{array}{cc}
M & m\\
m^{T} & 0_{3\times3}\end{array}\right),
\label{matrizse}
\end{equation}
with $M$ a $7\times 7$ matrix showing the mixing of neutralinos and right-handed neutrinos, and $m$ a $7\times 3$ matrix representing the mixing of neutralinos and right- and left-handed neutrinos. Both matrices can also be found 
in \cite{MuNuSSM, MuNuSSM2}.
The above $10\times 10$ matrix, Eq.~(\ref{matrizse}), is of the seesaw type giving rise 
to the neutrino masses which have to be very small. 
This is the case since the entries of the matrix $M$ are much
larger than the ones in the matrix $m$.
Notice in this respect that the entries of $M$ are of the order of the 
electroweak scale while the ones in $m$ are of the order 
of the Dirac masses for the neutrinos \cite{MuNuSSM,MuNuSSM2}.

Concerning the low-energy free parameters of the $\mu\nu$SSM in the neutral scalar sector, 
using the eight minimization conditions for the scalar potential,
one can eliminate the soft masses $m_{H_d}$, $m_{H_u}$, $m_{\widetilde{L}_{i}}$, and $m_{\widetilde{\nu}_{i}^{c}}$
in favour of the VEVs
$v_d$, $v_u$,
$\nu_i$, and $\nu^c_i$.
On the other hand, using the Standard Model Higgs VEV,
$v\approx 174$ GeV, 
$\tan\beta$, and $\nu_i$, one can determine the SUSY Higgs 
VEVs, $v_d$ and $v_u$, through $v^2 = v_d^2 + v_u^2 + \nu_i^2$. 
We thus consider as independent parameters
the following set of variables:
\begin{equation}
\lambda, \, \kappa,\, \tan\beta, \, \nu_1, \,  \nu_3, \nu^c, \, A_{\lambda}, \, A_{\kappa}, \, A_\nu\ ,
\label{freeparameters2}
\end{equation}
where we have assumed for simplicity that there is no intergenerational mixing
in the parameters of the model,
and that in general they have
the same values for the three families:
${\lambda} \equiv {\lambda_i}$, 
$\kappa\equiv {\kappa_{iii}}$, 
$\nu^c \equiv \nu^c_i$, $A_{\lambda} \equiv A_{\lambda_i}$, 
$A_{\kappa} \equiv A_{\kappa_{iii}}$,
$A_{\nu} \equiv A_{\nu_{ii}}$.
In the case of neutrino parameters, 
following the discussion in \cite{neutrinos,MuNuSSM2}, 
we need at least two generations with different VEVs and couplings in order to obtain the
correct experimental pattern. Thus we have choosen 
$Y_{\nu_1} \neq Y_{\nu_2}= Y_{\nu_3}$ and $\nu_1\neq \nu_2=\nu_3$.

The soft SUSY-breaking terms, namely gaugino masses,
$M_{1,2,3}$, scalar masses, $m_{\tilde Q,\tilde u^c,\tilde d^c,\tilde e^c}$,
and trilinear parameters, $A_{u,d,e}$, are also
taken as free parameters and specified at low scale.

\section{Gravitino dark matter}

Let us now show that the lifetime of the gravitino LSP is typically much longer than
the age of the Universe in the $\mu\nu$SSM, and therefore it can be in principle a candidate for DM.
In the supergravity Lagrangian there is an interaction term between
the gravitino, the field strength for the photon, and the photino. 
Since, as discussed above, due to the breaking of R-parity the photino and the left-handed neutrinos are mixed, the gravitino will be able to decay through the
interaction term into a photon
and a neutrino \cite{Takayama:2000uz}.
Thus one obtains:
\begin{equation}
\Gamma(\Psi_{3/2}
\to \sum_i \gamma \nu_i)\simeq  \frac{1}{32 \pi} \mid U_{\widetilde{\gamma} \nu} \mid^2 \frac{m^3_{3/2}}{M_P^2} \label{c1}
\ ,
\end{equation}
where $m_{3/2}$ is the gravitino mass, $M_P=2.4\times 10^{18} \gev$ is the reduced Planck mass, and 
$|U_{\widetilde{\gamma}\nu}|^{2}$ determines the photino content of the neutrino
\begin{equation}
|U_{\widetilde{\gamma}\nu}|^{2}=\sum_{i=1}^{3}|N_{i1}\cos\theta_{W}+
N_{i2}\sin\theta_{W}|^{2}\ .
\end{equation}
Here $N_{i1}$ ($N_{i2}$) is the Bino (Wino) component of the $i$-neutrino.

The lifetime of the gravitino can then be written as
\begin{equation}
\tau_{3/2} \simeq 3.8\times 10^{27}\ {s}
\left( \frac{\left| U_{\photino \nu}  \right|^2}{10^{-16}} \right)^{-1}
\left(\frac{m_{3/2}}{10\ \GeV} \right)^{-3}.
\label{lifetime25}
\end{equation}
If
$|U_{\widetilde{\gamma}\nu}|^{2}\sim 10^{-16}-10^{-12}$ in order to reproduce neutrino masses, as we will show below, the gravitino will be very long lived 
as expected (recall that the lifetime
of the Universe is about $10^{17}$ s).

For the gravitino to be a good DM candidate we still need to check that it can be present in the right amount to explain the relic density
inferred by WMAP, $\Omega_{DM} h^2 \simeq 0.1$.
As it is well known, adjusting the
reheating temperature after the inflatinary period of the Universe, one can reproduce the correct relic density for each possible value of the gravitino mass.
For example for $m_{3/2}$ of the order of $1-1000$ GeV, as expected 
from supergravity scenarios, one obtains $\Omega_{3/2} h^2 \simeq 0.1$ for
$T_R\sim 10^8-10^{11}$ GeV, with gluino masses of order $1$ TeV.
It is worth noticing here that
even with a high value of $T_R$ there is no gravitino problem, since the
next-to-LSP decays to standard model particles much earlier than BBN epoch
via R-parity breaking interactions.

Let us now show that
$|U_{\widetilde{\gamma}\nu}|^{2}\sim 10^{-16}-10^{-12}$ in the $\mu\nu$SSM.
We can easily make an estimation \cite{recentgravitino}. For a $2\times2$ matrix,
\begin{equation}
\left(\begin{array}{cc}
a & c\\
c & b\end{array}\right),
\end{equation}
the mixing angle is given by $\tan2\theta=2c/(a-b)$.
In our case (see Eq.~(\ref{matrizse}) and Refs. \cite{MuNuSSM, MuNuSSM2}) $c\sim g_{1}\nu\sim 10^{-4}$ GeV (represents the mixing of Bino and left handed
neutrino), $a\sim1$ TeV (represents the Bino mass $M_{1}$), and $b=0$.
Thus one obtains $\tan2\theta\sim 10^{-7}$, implying
$\sin\theta\sim\theta\sim10^{-7}$.
This gives $|U_{\widetilde{\gamma}\nu}|^{2}\sim10^{-14}$.
More general, $\theta\sim\frac{g_{1}\nu}{M_{1}}\sim10^{-6}-10^{-8}$, giving rise to
\begin{equation}
10^{-16} \lsim |U_{\widetilde{\gamma}\nu}|^{2} \lsim 10^{-12} \;.
\label{representative}
\end{equation}
We have carried out the numerical analysis of the whole parameter space 
discussed in Sect. 2, and the results confirm this estimation \cite{recentgravitino}.

\section{Gamma rays from gravitino decay}

Since in R-parity breaking models the gravitino decays producing a monochromatic photon with an energy $m_{3/2}/2$, one can try to extract constraints on the parameter space
from gamma-ray observations~\cite{Takayama:2000uz}.
Actually, model independent constraints on late DM decays
using the gamma rays were studied in~\cite{Yuksel:2007dr}.
\begin{figure}[!t]
  \begin{tabular}{c c}
   \includegraphics[width=0.5\textwidth]{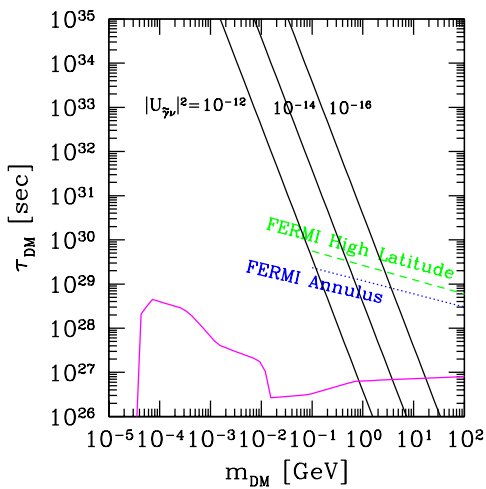} 
&
   \end{tabular}
 \caption{Constraints on lifetime versus mass for
a decaying DM particle, as explained in the text.}
\label{ExclusionPlot}
\end{figure}
There, the decaying DM was constrained using the gamma-ray line emission limits from the galactic center region obtained with the SPI spectrometer 
on INTEGRAL satellite, and the isotropic diffuse photon background as determined from SPI, COMPTEL and EGRET data.
These constraints are shown in Fig.~\ref{ExclusionPlot}
(from Ref. \cite{recentgravitino}), 
where the region below the magenta line is excluded. A
conservative non-singular profile at the galactic center is used.

On the other hand, the FERMI satellite~\cite{Fermi0} launched in June 2008 is able to 
measure gamma rays with energies between 0.1 and 300 GeV.
We also show in Fig.~\ref{ExclusionPlot} 
the detectability of FERMI 
in the 'annulus' and 'high latitude' regions
following the work in~\cite{bertone}.
Below the lines, FERMI will be able to detect the signal from decaying DM.
Obviously, no signal means that the region would be excluded and FERMI would have
been used to constrain the decay of DM~\cite{bertone}.

Finally, we show in the figure with black solid lines the values of the parameters predicted by the $\mu\nu$SSM using Eq.~(\ref{lifetime25}), for
several representative values of 
$|U_{\widetilde{\gamma}\nu}|^{2}$ discussed in Eq. (\ref{representative}).
We can see that values of the gravitino mass larger than 20 GeV are disfavored in this model by the isotropic diffuse photon background observations 
(magenta line).
In addition, FERMI will be able to check important regions of the parameter 
space with gravitino mass between $0.1 - 20$ GeV and $|U_{\widetilde{\gamma}\nu}|^{2}=10^{-16}-10^{-12}$ (those below the green line).

Let us now discuss in more detail \cite{recentgravitino} what kind of signal is expected to be observed by FERMI
if the gravitino lifetime and mass in the $\mu\nu$SSM (black solid lines) correspond to a point below the
green line 
in Fig.~\ref{ExclusionPlot}

As it is well known, there are two sources for a diffuse background from DM decay.
One is the cosmological diffuse gamma ray coming from extragalactic regions, and the other is the one coming from the halo of our galaxy.

The photons from cosmological distances are red-shifted during their journey to 
the observer and the isotropic extragalactic flux can be found in \cite{Takayama:2000uz,bertone}.
On the other hand, 
the photon flux from the galactic halo shows an anisotropic sharp line.
For decaying DM this is given by
\begin{equation}
\frac{d J_{halo}}{dE}= A_{halo}\frac{2}{\mdm}\delta\left(1-\frac{2E}{m_{DM}} \right)
\;\;\;;\;\;\;
A_{halo}= \frac{1}{4\pi\taudm\mdm}\int_{\textrm{los}}\rho_{halo}(\vec{l})d\vec{l}\ ,
\end{equation}
where the halo DM density is integrated along the line of sight, and
we will use a NFW profile, 
$\rho_{NFW}(r)=\frac{\rho_h}{r/r_c(1+r/r_c)^2}$,
where we take $\rho_h=0.33 \gev/\cm^3$,
$r_c=20$ kpc, and
$r$ is the distance from the center of the galaxy.
The latter 
can be re-expressed using the distance from the Sun, $s$, in
units of $\Rsun=8.5$ kpc (the distance between the Sun and the galactic center) and
the galactic coordinates, the longitude, $l$, and the latitude, $b$, as
$r^2(s,b,l)=\Rsun^2[(s-\cos b\cos l)^2+(1-\cos^2b \cos^2l)]$.

\begin{figure}[!t]
  \begin{tabular}{c c}
   \includegraphics[width=0.5\textwidth]{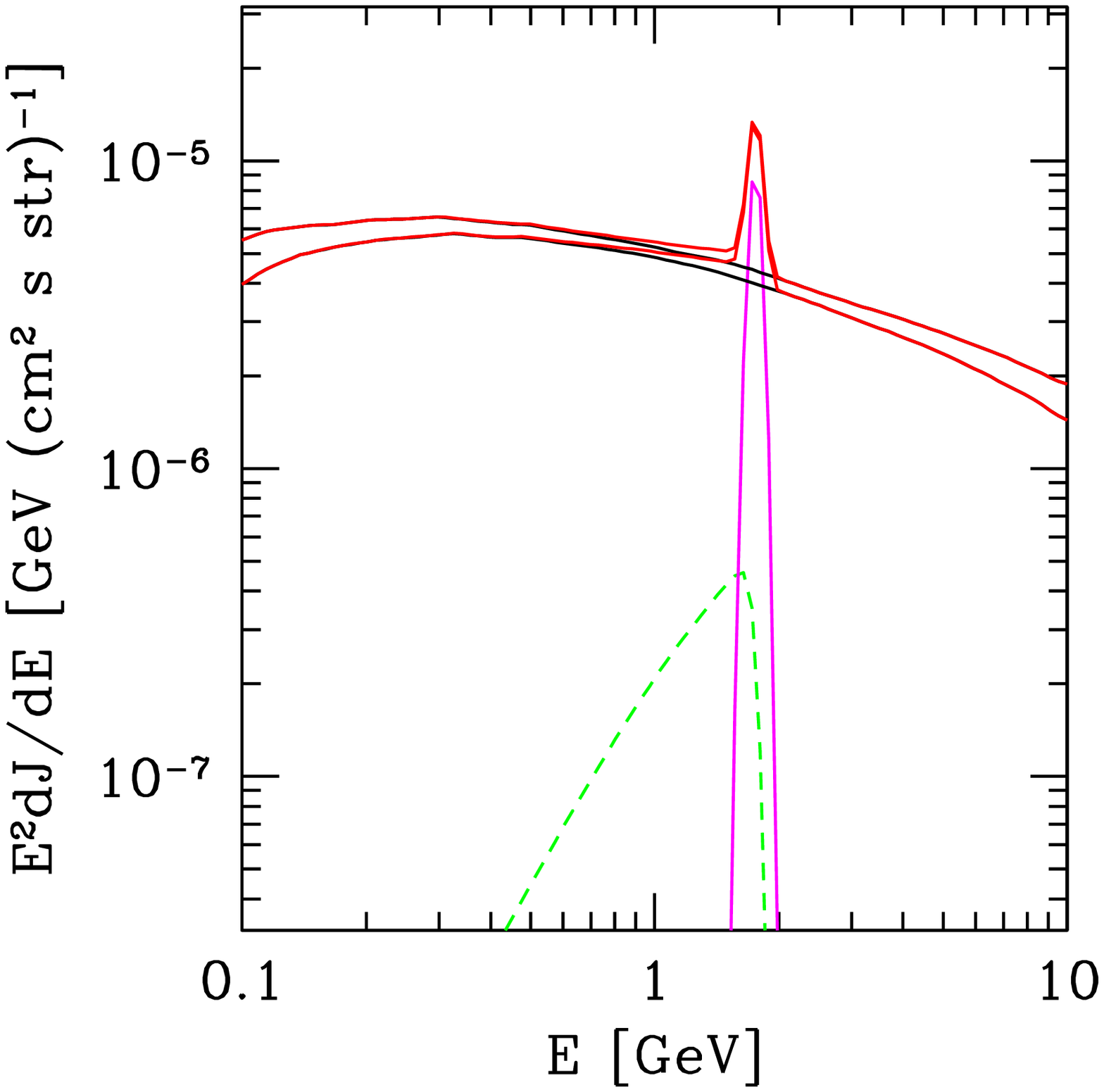}
&
   \includegraphics[width=0.5\textwidth]{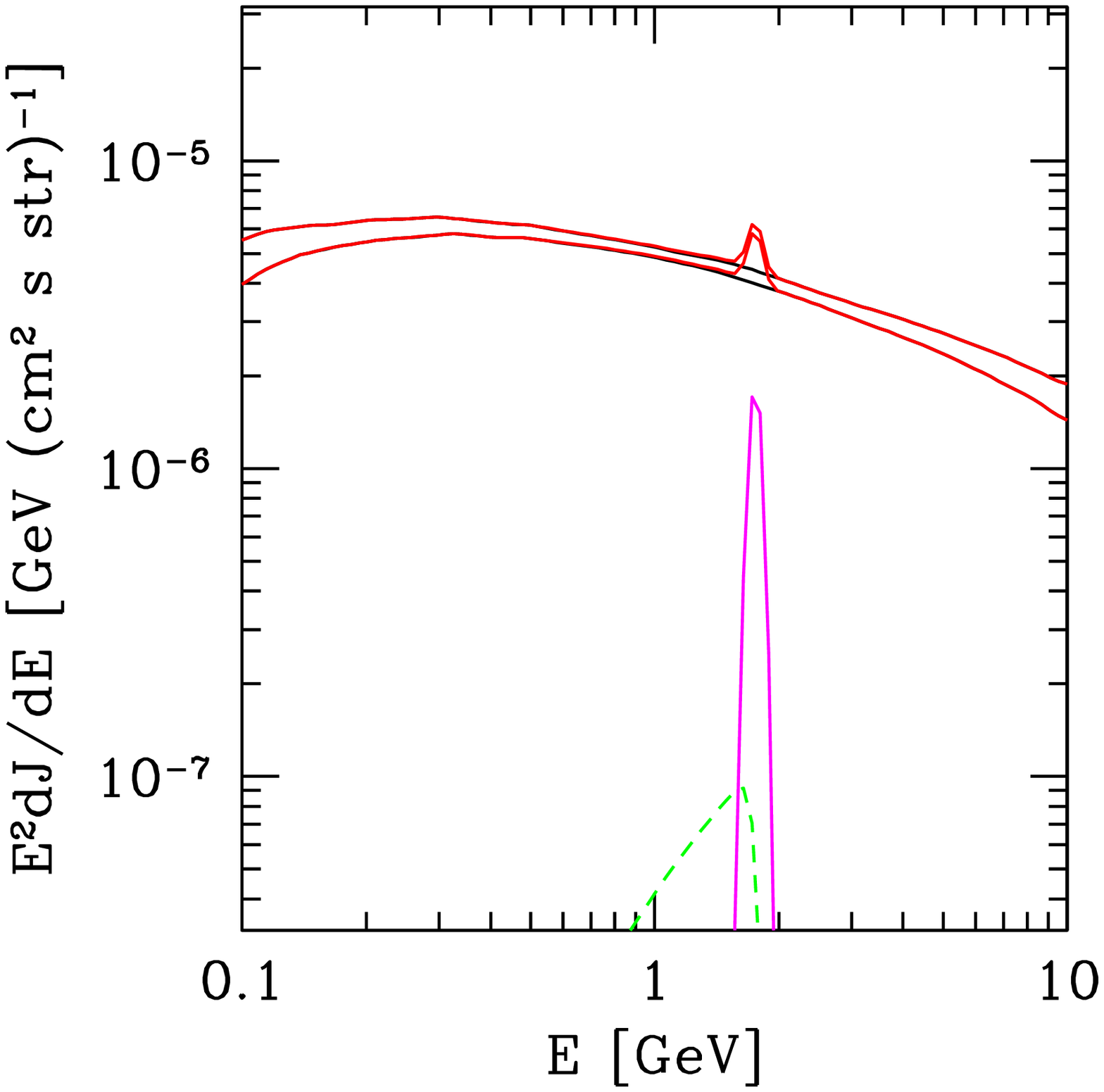}
      \\ (a)& (b)\\
   \end{tabular}
 \caption{
Expected gamma-ray spectrum 
for an example of gravitino DM decay 
in the mid-latitude range 
in the $\mu\nu$SSM
with 
$m_{3/2}=3.5\gev$ and (a)
$\left| U_{\photino \nu}  \right|^2= 8.8\times10^{-15}$, 
(b) 
$\left| U_{\photino \nu}  \right|^2= 1.7\times10^{-15}$.
} \label{Gravitinodecay}
\end{figure}

As an example, let us compute with these formulae the expected diffuse gamma-ray
emission in the mid-latitude range ($10^{\, \circ}\le |b|\le 20^{\,\circ}$),
which is being analized by FERMI,
for the case of gravitino DM.
Let us assume for instance a value of $m_{3/2}=3.5$ GeV and
$\left| U_{\photino \nu}  \right|^2= 8.8\times10^{-15}\, (1.7\times10^{-15})$ in the $\mu\nu$SSM, corresponding to
$\tau_{3/2}=10^{27}$ ($5\times10^{27}$) s, using Fig.~\ref{ExclusionPlot}.
We convolve the signal with a Gaussian distribution with the energy resolution  $\Delta E/E=0.09$, between $E=1-10\gev$, following~\cite{Fermi0}, and  
then we average the halo signal over the region for the mid-latitude range mentioned above.

The results for the two examples are shown in Fig.~\ref{Gravitinodecay}, where the green dashed line corresponds to the diffuse extragalactic gamma ray flux,
the magenta solid line corresponds to the gamma-ray flux from the halo, and
the black solid lines represent the conventional background.
The total gamma-ray flux,
including background, extragalactic, and line signal, is shown with red solid lines.
We can see that
the sharp line signal associated to an energy half of the gravitino mass, dominates the extragalactic signal
and can be a direct measurement (or exclusion) in the FERMI gamma ray observation.

\begin{theacknowledgments}
I gratefully acknowledge the local organizers of DSU 09 for the
wonderful atmosphere that they created.
I also thank one of the participants, A. Morselli, for very helpful information concerning
FERMI.
This work was supported 
in part by the Spanish MICINN under grants FPA2006-01105 and FPA2006-05423,
by the Comunidad de Madrid under grant HEPHACOS P-ESP-00346,
and by the European Union under the RTN program MRTN-CT-2004-503369. 
We also
thank the ENTApP Network of the ILIAS project RII3-CT-2004-506222 
and the UniverseNet Network MRTN-CT-2006-035863.
\end{theacknowledgments}



\bibliographystyle{aipproc}   

\bibliography{sample}

\IfFileExists{\jobname.bbl}{}
 {\typeout{}
  \typeout{******************************************}
  \typeout{** Please run "bibtex \jobname" to optain}
  \typeout{** the bibliography and then re-run LaTeX}
  \typeout{** twice to fix the references!}
  \typeout{******************************************}
  \typeout{}
 }


\end{document}